\documentclass[]{aa}
\usepackage{graphics}
\usepackage{psfig}
\usepackage{balance}
\begin{document}
\def\teff{$T\rm_{eff }$ }
\def\logg {log\,g}
\def\lambo{$\lambda$ Boo }
\def\vsini {$v\sin i$ }
\def\kms {$km\, s^{-1}$ }
\title{  
\lambo stars  with composite spectra 
\thanks{Based on
observations collected at ESO (Echelec spectrograph) and at TBL of the Pic
du Midi Observatory (France) 
 }}
   \subtitle{}
\author{R. \,Faraggiana \inst{1}
\and P. Bonifacio \inst{2}
\and E. Caffau \inst{2}
\and M. Gerbaldi \inst{3,4}
\and M. Nonino \inst{2}
}
  \offprints{R. Faraggiana}

\institute{
Dipartimento di Astronomia, Universit\`a degli Studi di Trieste,
Via G.B.Tiepolo 11, I-34131 Trieste, Italy \\
email: faraggiana@ts.astro.it
\and
Istituto Nazionale per l'Astrofisica --
Osservatorio Astronomico di Trieste,
Via G.B.Tiepolo 11, I-34131 Trieste, Italy 
\and
Institut d'Astrophysique, 98 bis Bd Arago, F-75014 Paris, France 
\and
Universit\'e de Paris Sud XI
}

\mail{faraggiana@ts.astro.it}
\authorrunning{R. Faraggiana et al. }
\titlerunning{\lambo stars with composite spectra}

\date{Received ... / Accepted ...}
\abstract{ 
We examine the large sample of \lambo candidates collected in 
Table 1 of Gerbaldi et al. (2003) to see how many of them show
composite spectra.
Of the 132 \lambo candidates we identify 22 which 
definitely show composite spectra and
15 more for which there are good reasons
to suspect a composite spectrum.
The percentage of \lambo candidates with
composite spectra is therefore $> 17$ and possibly
considerably higher.
For such stars the \lambo classification
should be reconsidered taking into account the
fact that their spectra are composite.
We argue that some of the underabundances
reported in the literature may simply be
the result of the failure to consider the
composite nature of the spectra.
This leads to the legitimate suspicion
that some, if not all, the \lambo candidates 
are not chemically peculiar at all.
A thorough analysis of even
a single  one of the \lambo candidates
with composite spectra, in which the composite
nature of the spectrum is duly considered,
which would demonstrate that the chemical
peculiarities persist, would clear 
the doubt we presently have that   the 
stars with composite spectra
may not  be \lambo at all.

}
\maketitle{}

\section{Introduction}

The \lambo stars 
are a fascinating class of stars to which much
attention has been devoted in the recent years.
It is the  only class of A-type stars with abundances lower
than solar, which however have all the kinematical
and photometric properties of Pop I stars. 
It is widely accepted that their low metallicity
cannot be ascribed to an age effect and is in fact
unrelated to the chemical evolution of the Galaxy. 

Several
unsuccessful attempts have been made in the past to interpret this
peculiar behaviour from the first
historical theory of spallation reactions (Sargent, 1965)
to the Michaud \& Charland (1986) theory of diffusion/mass-loss
the  weakness of which 
has been underlined by Baschek \& Slettebak (1988) 
and examined in detail by
Charbonneau (1993). These 
authors raised the critical problem
of the thermally-driven meridional circulation generated by the
relatively rapid, when
compared to that of Am stars, rotation of these stars. 
In \lambo stars
a vigorous
meridional circulation should  prevent diffusion to operate efficiently.

The best theory,  at present, is based on
the diffusion-accretion model proposed by Venn \& Lambert (1990)
to explain the abundance anomalies of the three \lambo stars studied by them.
This hypothesis has been refined later on by Waters et al. 
(1992) and by Turcotte \& Charbonneau (1993).
The key point of these models is the existence of a surrounding disk of dust 
and gas which would be the remnant of the star formation material. 
The gaseous part would be accreted by the star, while the dust would be blown
away by radiation pressure. According to this theory  the \lambo stars should 
be young objects,
in the last phases of their pre-main sequence life or in the early phases
of the main-sequence. In fact it has been shown that any remnant of the
initial disk will be blown out in 10$^6$ yrs upon termination of accretion
episode. 
It is in fact true that 
some \lambo stars  
have IR excess and/or shell signatures which could be explained
by circumstellar material still surrounding a very young star or system.
Also  \lambo have been found to be non-radial pulsating objects,
which would agree with the hypothesis that they lie in the zone of pulsation
instability.

The hypothesis that the 
\lambo stars
may be young objects, has been explored by several authors since the first
study by Gerbaldi et al. (1993), but it turned out not 
to be a general explanation for the \lambo stars. In fact
it appears that they   occupy a very  large
domain of the HR diagram.

A similar model in essence, but with an alternative scenario has been
proposed by Kamp \& Paunzen (2002) where the surrounding 
disk is replaced by a diffuse IS cloud crossed by the star.
In such a model it is however unclear why such
effects should be limited to the \lambo
range of $T\rm_{eff }$.

The objects classified as \lambo display a  large variety 
of properties 
and a remarkable lack of any 
relation between their physical parameters ($T\rm_{eff }$, log g, $v\sin i$, age, 
galactic coordinates) and the measured abundance anomalies.
Recently  bright companions were detected  near several \lambo candidates
by ground-based speckle and adaptive optics (AO)
observations, as well as  by the Hipparcos 
space experiment.
This situation prompted us to explore  a new hypothesis 
to explain the \lambo phenomenon:
a combination of two similar
spectra can be confused with that of a single metal poor star  
(Faraggiana \& Bonifacio, 1999).

In previous papers on this subject we showed that composite spectra due to
undetected binary stars may be easily confused with those of a single
peculiar star classified as $\lambda$ Boo. 
In particular, from a detailed inspection 
of high resolution  spectra of close binaries producing composite spectra, 
but classified as single peculiar stars belonging to  \lambo class, we 
have selected some criteria that allow to distinguish
composite spectra (Faraggiana et al., 2001a) .

These criteria have been successfully used to demonstrate that other
\lambo stars are in fact  binaries which give rise to  composite spectra
(Faraggiana et al., 2001a; Faraggiana et al., 2001b; Faraggiana \& 
Gerbaldi, 2003).  More recently we 
have shown that a high
percentage of \lambo candidates have visual and UV photometric properties 
(Gerbaldi et al., 2003) that are incompatible with the \lambo classification.

We are continuing our research aiming to select a group of spectra 
of metal-deficient Pop I A type non-contaminated by binarity.

In the present paper we recall in section 2 the previously defined 
binary detection criteria and clarify how they are often complementary;
we discuss, in section 3, the new \lambo candidates which appear to be 
binaries with the spectrum contaminated by that of a companion;
in section 4 we  discuss 
the \lambo candidates with composite spectra
and in the conclusions we examine the possible relationship, if any, between 
duplicity and \lambo phenomenon.

\section{Binary stars producing composite spectra}

\subsection{Duplicity detection by  imaging: Hipparcos, speckle 
interferometry, adaptive optics}

Binary systems detected  by imaging are discussed in detail by  
Gerbaldi et al. 
(2003). 
The measure of 
separation and magnitude difference allowed to recognize that 
12 objects, corresponding to 9\% of the total,
classified as $\lambda$ Boo, are in reality binaries that cannot be resolved at 
the  spectrograph
entrance. Their classification is based on the properties of the average 
spectrum of two stars of similar luminosity and spectral type; this 
combination in general produces a spectrum affected by veiling, so that the 
metal lines appear weak and may mimic those formed in a single metal deficient 
atmosphere.

\subsection{Duplicity detection from spectroscopic and photometric data} 

The detection of duplicity is tricky since these objects do not show,
in general, the classical double peaked narrow lines as in most classical
SB2. This is mainly due to the fact that one of the characteristics of \lambo 
stars is the mean-high \vsini which produces broad and shallow lines.
The detection of possible duplicity is therefore more promising in the 
visual-red
wavelength range where the blending is lower than at shorter wavelengths.

The OI triplet at 777.4 nm  is not affected by any other line; the shape of the
three components allows to distinguish the presence of 
a secondary set of lines if a companion of similar brightness at this 
wavelength is present; however this is true only up to 
\vsini $\simeq$ 100 km s$^{-1}$, value 
for which the 3 lines merge in one unique broad featureless
line. This very powerful criterion 
has been used to recognize spectroscopically
known binaries among 
the lower rotating \lambo candidates; an example is HD 47152 (Faraggiana 
et al., 2001a). 

The presence of a secondary component of very similar brightness and 
similar atmospheric parameters may be easily detected by  cross 
correlation  using as template a sharp lined synthetic spectrum 
(broadened to 5 km s$^{-1}$). An example is that of HD 11413
discussed in the next section. 
However, the cross correlation does not always reveal the presence of the 
companion star; this is the case when: 

\begin{enumerate}

\item the observed composite spectrum is detected 
by the same asymmetry of many lines (as in the case of HD 196821 discussed in 
section 3.4); 
\item the lines of the brighter component are much stronger than
those of the companion (as in HD 141851 discussed in section 3.3);
\item 
the chosen spectral range includes a strong Balmer line whose behaviour
dominates in the cross correlation 
\footnote{this last limitation can be overcome 
by suppressing
the Balmer line profile in both the observed and the template spectra, 
as it has been done for HD 11413, HD 79108 and HD 210111 discussed in 
next sections.}. 
\end{enumerate}

The core of Balmer lines should be deeper than that of the synthetic spectra
computed with the LTE approximation, as are those computed with  the
SYNTHE code (Kurucz, 1993)
used by us; a flat and square inner core profile is a sign of 
a probable duplicity (Faraggiana \& Bonifacio, 1999).
 
Other more subtle signatures of duplicity are the inconsistent results
obtained when abundance analysis is attempted. This
is what we have experienced, for example, 
with the known binary HD 47152, when we tried to fit its spectrum with that
computed by using the model based on the parameters derived from the
calibration of photometric colour indices. In this case the observed
Balmer line wings are roughly fitted by LTE computations, but metal lines,
in particular the many  Fe I and Fe II lines
do not respect the 
ionization equilibrium. Moreover the comparison with computations 
indicates that while some profiles are fitted by computations, for others
the computed lines appear to be either too strong or too weak. 

In some cases, subtle duplicity signatures can be detected when several 
spectra are available, because the effect of the companion is different at
different phases (e.g. HD 11413 and HD 210111 discussed in the 
following sections).

The presence of two sets of lines with different broadening as in the case of 
HD 111786 visual spectrum (Faraggiana et al., 1997) is also a sign of 
a composite 
spectrum. This fact, coupled with the absence 
of any narrow line component in the UV range (HR IUE spectra) allowed us
to deduce that the narrow lines are due to a secondary source cooler than
the primary. We point out that such 
narrow lines cannot be interpreted as due to 
a circumstellar shell, since   shell lines
have a different behaviour:  the presence of a shell produces
narrow components mainly of the resonance lines  
and of some low excitation lines and are not restricted to a limited
wavelength range.
This is the case of HD 38545 (Faraggiana et al., 2001a), for which
strong shell lines are seen both in the visual and in the UV spectrum; 
the binary nature of this object has been discovered by Hipparcos and,
on the basis of the spectra at our disposal,  
we have been unable to decide if the shell is surrounding one of the
binary system component or the whole system. In fact the broad photospheric
lines prevent the spectral detection of duplicity from spectrum inspection.

Large inconsistencies between the atmospheric parameters derived from
calibration of visual photometric colour indices and the UV flux distribution,
discussed in Gerbaldi et al. (2003) paper for the stars observed by TD1, 
are also signs 
that the star is not a classical $\lambda$ Boo.

\section{Detected composite spectra}

The target stars are extracted from the list of \lambo stars
compiled by Gerbaldi et al. (2003); this list contains all the stars
that have been classified as members of this class and for each of them 
the reference to the classifier(s) is given.

In this section we describe a few
stars for which we have collected data supporting
the interpretation as composite spectra.
The basic data for these stars  are given  in Table 1,
together 
with the observational data and some instrumental details.

The wavelength coverage of the spectra taken at ESO with 
the Echelec spectrograph is 421-450 nm 
since only the 9 central orders have been extracted. The nominal spectral
resolution of 70000 was degraded to 28000 by using a slit width
of 320 $\mu$ (1.5 arcsec on the sky). 
The reduction from the CCD images to the complete linear spectra 
(calibration frames, orders extraction, subtraction of background
intensity and scattered light, wavelength calibration 
and connection of the orders) has been done with a package 
which runs under MIDAS and was developed expressly for this 
spectrograph's data. The details of all the steps of the 
procedure are described in Burnage \& Gerbaldi (1990, 1992).

The spectra taken at TBL with the Musicos spectrograph cover the 
wavelength range 515-889 nm. This is a fiber-fed spectrograph
and its resolution is determined by the 
50 $\mu$ fiber corresponding to 2.1 arcsec on the sky
which yield R=32000.
The photometric and wavelength 
reductions have been made by using ESpRIT (Echelle spectra
Reduction: an Interactive Tool), a computer code for on-line
processing developed by Donati et al. (1997) and made available  
at the TBL telescope of the Pic du Midi.

The atmospheric parameters derived from the 
Str\"omgren and Geneva colour indices, as described in Gerbaldi et al. (2003),
are given in Table 2; these parameters are 
derived under the hypothesis that
the objects are single stars and  cannot be used for 
a reliable abundance 
determination. 

The synthetic spectra used in next sections have been computed by using 
the Kurucz model atmosphere grid and SYNTHE code (Kurucz, 1993).

\bigskip

\begin{table*} 
\caption{The new composite spectra: observational data}
\label{table_1}
\begin{center}
\begin{tabular}{rrcccccc}
\hline
\\ 
HD        & HR    & V  & tel & spectr. & R  & date & exp.time (mn) \\
11413     &  541  & 5.94 & 1.5 ESO & Echelec & 28000   & 16 Nov. 1992 &  50 \\
          &       &      &         &         &        & 09 Sep. 1993 &  50 \\ 
79108     & 3651  & 6.14 & 1.5 ESO & Echelec & 28000  & 15 Jan. 1995 &  55 \\ 
141851    & 5895  & 5.09 & 1.5 ESO & Echelec & 28000  & 04 Apr. 1993 &  33 \\   
          &       &      &         &         &        & 07 Apr. 1993 &  30 \\ 
          &       &      & 2.0 TBL & Musicos & 32000  & 20 Mar. 2000 &  60 \\
196821    & 7903  & 6.08 & 2.0 TBL & Musicos & 32000  & 06 Oct. 2002 &  120 \\ 
210111    & 8437  &   6.37 & 1.5 ESO & Echelec & 28000  & 07 Sep. 1993 &  120 \\   
          &       &        &         &         &        & 18 Sep. 1994 &  90 \\   
          &       &        &         &         &        & 14 Nov. 1994 &  55 \\   
\\
149303    & 6162  & 5.65& 2.0 TBL    & Musicos & 32000& 17 Feb. 2003 &60\\
\\

\hline
\end{tabular}
\\
\end{center}
\end{table*}

\bigskip

\begin{table}
\caption{The new composite spectrum stars: atmospheric parameters derived
from photometry}
\label{table_2}
\begin{center}
\begin{tabular}{rcccccc}
\hline
\\
HD        & E(b-y)    & \teff   & log g   & \teff  & log g   & [M/H]   \\
          &  MD      &   MD     & MD     &   Gen  &   Gen   &   Gen  \\
11413     & 0.002  &    7950 &  3.84 &     7813 &   4.08  & -2.03     \\
79108     &  0.004  &    9830 &  4.07  &    9852 &   4.17  &  --     \\
141851    & -0.014  &    8080 &  3.86  &    8258 &   3.69  &  --      \\
196821    & 0.002  &   10390 &  3.55  &   10260 &   3.63  &  --      \\
210111    & -0.023  &    7450 &  3.75  &    7545 &   4.26  & -1.18       \\
\\
149303    & -0.018 &    8120 & 3.80 &   8483     &3.78    &  --      \\
\\
\\
\hline
\end{tabular}
\\
\end{center}
\end{table}

The validity of Moon and Dworetsky (1985) (MD) and K\"unzli et al. (1997) 
(Gen) calibrations for metal deficient 
stars may be questioned, however we believe 
it is amply sufficient for the present purpose.

\subsection{HD 11413}

This is one of the few stars unanimously classified as \lambo 
(Abt \& Morrell, 1995; Gray, 1988; Paunzen et al., 1997; Paunzen, 2001).
Abundances for this star have been discussed in several papers
and are all based on the hypothesis that this is a single star.
The star is an unsolved variable (U) according to Hipparcos data 
and variable of $\delta$ Sct type according to Adelman et al. (2000).
The spectra used in the present paper have been already used by Grenier 
et al. (1999) for precise 
RV measures and the star was  classified in that paper as probable double.
We made a finer check for duplicity correlating the spectra with the
synthetic spectrum computed  
with the parameters obtained from uvby$\beta$ photometry calibration and a 
low value of \vsini (5 \kms).
Due to the predominant  r\^ole of the H$_{\gamma}$ line in the 300 \AA~ range
covered by our spectra and to the possible inaccuracy of the continuum
drawing through such a broad line in dwarf A-type stars which covers 3 
orders of the echelle spectrograph, we  artificially
suppress H$_\gamma$ by  normalizing with a cubic spline
through the line profile, both on observed and
computed spectra before performing the cross-correlation.

The cross correlation program rebins the observed and the template
in the velocity space, the rebinning is largely oversampled; the 
correlation index is normalized to 1 and in the abscissa the RV is given.

\begin{figure}
\psfig{figure=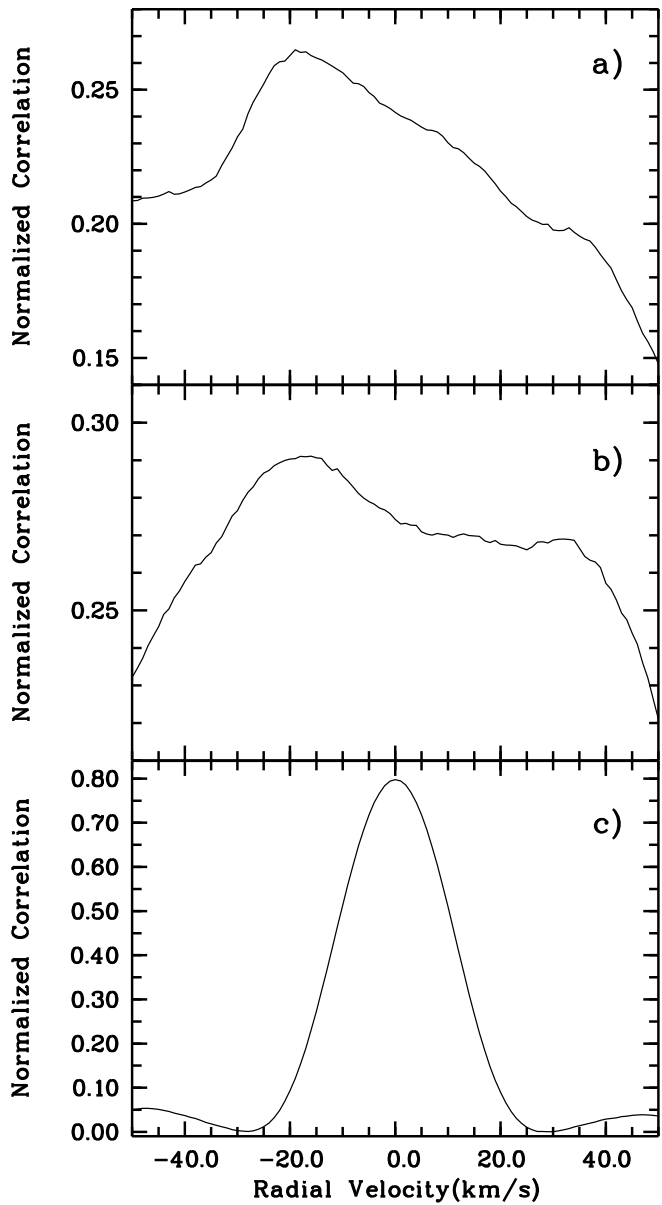,clip=true}
\caption{ 
Panel a):the cross correlation of the spectrum of
HD 11413 taken on Nov 16th, 1992 with a computed
spectrum  (\teff = 7950K, log g = 3.84, $v\sin i=5$ \kms)
over the range 420-450nm. In both spectra the 
 H$_\gamma$ has been artificially suppressed (see text).
Panel b):The cross correlation of  the spectrum of HD 11413
taken on Sep. 9th, 1993 with the same computed spectrum 
as used in panel b); also the spectral range is the same
and H$_\gamma$ has been likewise suppressed.
Panel c): the cross correlation of  the spectrum of Sirius;  
with a (\teff = 9830 K, log g = 4.07, $v\sin i=5$ \kms)
suitable synthetic spectrum,
the spectral range is the same
and H$_\gamma$ has been  suppressed as for HD 11413.
}
\label{hd11413_cor}
\end{figure}

The cross correlations (Fig. \ref{hd11413_cor} ) indicate clearly
that the spectrum is formed by more than a 
single source. This is more evident for the HD 11413 spectrum taken on 1993,
Sep 9th, where a separation of about
48 \kms between the two components has been measured. 
The existence of a third body may be suspected from the shape of 
the cross correlation of the spectrum taken on 1992, Nov 16th (left wing).
   
In order to allow the reader to see what the
cross-correlation of a star with a non-composite
spectrum looks like,
in the bottom panel of Fig. \ref{hd11413_cor} we show 
the cross correlation of an H$_{\gamma}$ suppressed  spectrum
of Sirius, with an appropriate
synthetic spectrum. 
The normalized peak is symmetric and its height is almost
0.8. 
Sirius is a known binary in which the secondary is a white
dwarf, therefore much fainter and its spectrum is not
composite. The difference between this cross correlation
and those of HD 11413 is striking. 

HD 11413 belongs to the group of pulsating \lambo and the detailed 
analysis of its pulsation characteristics has been made by Koen 
et al. (2003); this allows to study the impact of pulsations on the 
spectra analyzed here.
These authors have shown that no line profile changes have been 
detected in the 3 days covered by their observations made with 
R=39000, higher than that of the spectra used by us.
Having taken a large series of spectra with very short exposures
(240 to 500 s) these authors have been able to measure the RV 
variations connected with pulsations. Figure 2 of Koen et al. shows
that the amplitude of these RV is 3 \kms.

The RV difference between the two peaks found in
the correlation of our spectrum  of HD 11413 taken on
September 9th 1993
is almost one order of magnitude higher. 
It cannot be due to the star pulsation; we also recall that
our spectra, taken with 50m of exposure, average the complex pulsation
period (about 0.04 d), so reducing the visibility of NRP.

The comparison between observations and computations shows a poor
agreement with the synthetic spectra computed both with the parameters 
derived from uvby$\beta$ or Geneva photometries.

The \vsini derived by Royer et al. (2001) from the frequency
of the first zero of the Fourier transform of 3 lines 
of these same spectra is 140 $\mathrm{km\, s^{-1}}$, 
a value slightly higher than those measured by St\"urenburg (1993) 
(125 $\mathrm{km\, s^{-1}}$ )
and by Holweger \& Rentzsch-Holm (1995) (122 $\mathrm{km\, s^{-1}}$).
These latter authors derived \vsini from the CaII K line profile;
the discrepancy between this value and that derived from the
metal lines in the 420--450 nm region is likely due 
to the fact that it is almost impossible to normalize correctly
the observed spectrum in this region where the overlap of H$_\epsilon$
and H$_8$ broad wings prevent to see the real continuum.

So HD 11413 is a new SB2  or multiple star  
and it is another example of peculiar hydrogen line (PHL), common
among \lambo stars, which is explained by the duplicity of the
object.

\subsection {HD 79108}

\begin{figure}
\psfig{figure=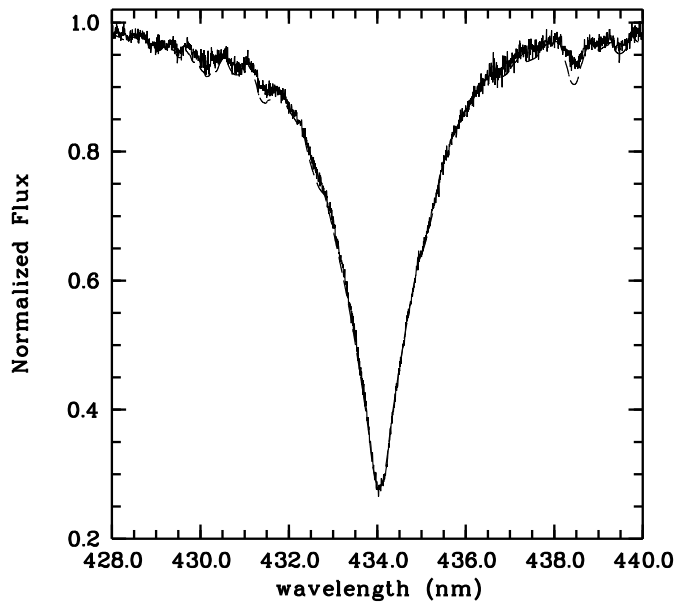,width=8.8cm,angle=0,clip=true}
\caption{ The H$_{\gamma}$ profile of HD 79108 compared 
with a spectrum computed with \teff = 9830 K, log = 4.07 and
\vsini=160 km s$^{-1}$.}
\label{hgamma79108}
\end{figure}

\begin{figure}
\psfig{figure=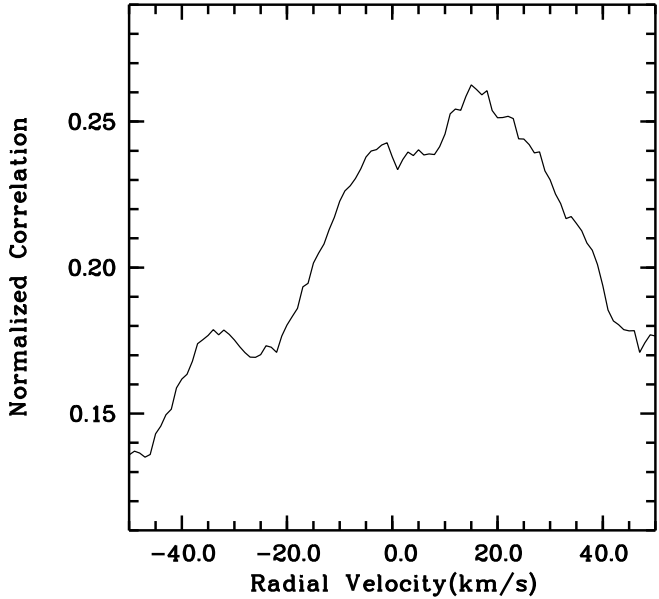,clip=true}
\caption{ The cross correlation of the spectrum of
HD 79108 with the synthetic spectrum computed with  \teff= 9830 K, log = 4.07 and
$v\sin i=5$ \kms and with 
H$_\gamma$ suppressed (see text).}
\label{fig_hd79108}
\end{figure}

The star is a \lambo candidate having a slightly negative $\Delta$a index
(Maitzen \& Pavlovski, 1989; Vogt et al., 1998); it has also been classified
\lambo by Abt \& Morrell (1995). King (1994) has derived an IR excess 
at 60 $\mu$m from
IRAS data. Grenier et al. (1999) classified it as suspected double,
but speckle interferometric observations (Ebersberger et al., 1986) gave a 
negative result. Its RV is variable according to the Bright Star Catalog 
in spite of the fact 
that the star does not belong to a known binary system.
The H$_\gamma$ profile is fitted by the spectrum computed with the MD 
parameters (Fig. \ref{hgamma79108}); however the observed core is not deeper 
than the computed one,
as expected (see section 2.2). The metal lines have square and asymmetric
profiles  suggesting a composite spectrum.
The cross correlation made without H$_\gamma$, as for HD 11413, confirms
that the object is in reality a complex system composed by at least two stars
of similar luminosity and a third less luminous component 
(Fig. \ref{fig_hd79108}).

\subsection {HD 141851}

This star has been classified \lambo candidate by  Abt  (1984) 
on the basis of
the weak MgII 448.1nm line; the colour indices in the Geneva photometric
system support this classification (Hauck, 1986).
The star is included in the Paunzen et al. (1997) "consolidated catalogue of 
\lambo stars", but not in the list of "confirmed members" of this group
by Paunzen (2001).
 The star is  a known binary,
first found by McAlister et al. (1987) through
their speckle measures; the separation varies from 0.069 to 0.132 arcsec
(Hartkopf et al., 2003), but no estimate of the 
magnitude difference is given by these authors.
However, the orbital period seems to be very long, probably of more than 60 yr, 
according to Iliev et al. (2001).

According to the spectrum analyzed by Grenier et al. (1999) the star belongs
to the group of certain double A-F type with faint F-G component, i.e.
the observed spectrum appears to be  significantly contaminated by that of
the fainter and cooler companion.
AO observations (Gerbaldi et al., 2003) have demonstrated
that the flux of the secondary star, is about 1/3 of that of the primary
in the H filter and cannot be neglected.

The \vsini values found in the literature are quite
diverse: 185 \kms (Abt \& Morrell, 1995),  230 \kms (Royer, 2001),
200 \kms for C lines 
and 280 \kms for O lines (Paunzen et al., 1999a), 
260 \kms (Andrievsky et al., 2002). 
None 
of these authors mentions the presence of 
weak narrow lines all along the spectrum.

This star belongs
to the sample of \lambo for which Paunzen et al. (1999a) and Kamp et al. (2001)
determined NLTE abundances of C, O and Ca in LTE  respectively by using the
average parameters of the two system components. 
In 2002 Andrievsky et al.
made a new LTE abundance determination of C, O, Si and Fe; they adopted the 
same model as the previous authors. 
Abundance differences
with the results obtained by the previous authors appears for C and O.
In Paunzen et al (1999a) the given LTE abundance of C is -0.73 (NLTE=-0.81)
to be compared to the value by Andrievsky et al. : -0.20; in the same paper
the LTE abundance of O is +0.25 (NLTE=-0.21)
to be compared to the value by Andrievsky et al : -0.38.

The LTE abundance difference of C is considerably  larger than what a 
state of the 
art abundance analysis should provide; they are based on different lines, 
so probably differently affected by NLTE. The wavelength of the lines used 
by  Andrievsky et al
is not specified so no abundance correction for NLTE can be estimated.
Another  possible interpretation of these different abundances is that the
two groups used spectra taken at different dates, so probably
at different phases of the binary system. 

Hauck et al. (1998) measured the EW (2.3 m\AA) of a narrow component in
the core of the K-line; they interpreted it as a signature of circumstellar
matter around the star. This narrow feature may be interpreted more simply
as the the spectral signature of the cooler companion.

An X-Ray emission, unexpected in early-, middle-A-type stars has been measured
by ROSAT PSPC, and correctly ascribed to the companion star
(Simon et al. 1995).
In fact, if one component of this system has a T$_{eff}$ lower than
8000 K, and thus a convective envelope from which X-Rays may originate,
the detection of the X-Ray emission from the system does not
present any peculiarity.

\begin{figure*}
\psfig{figure=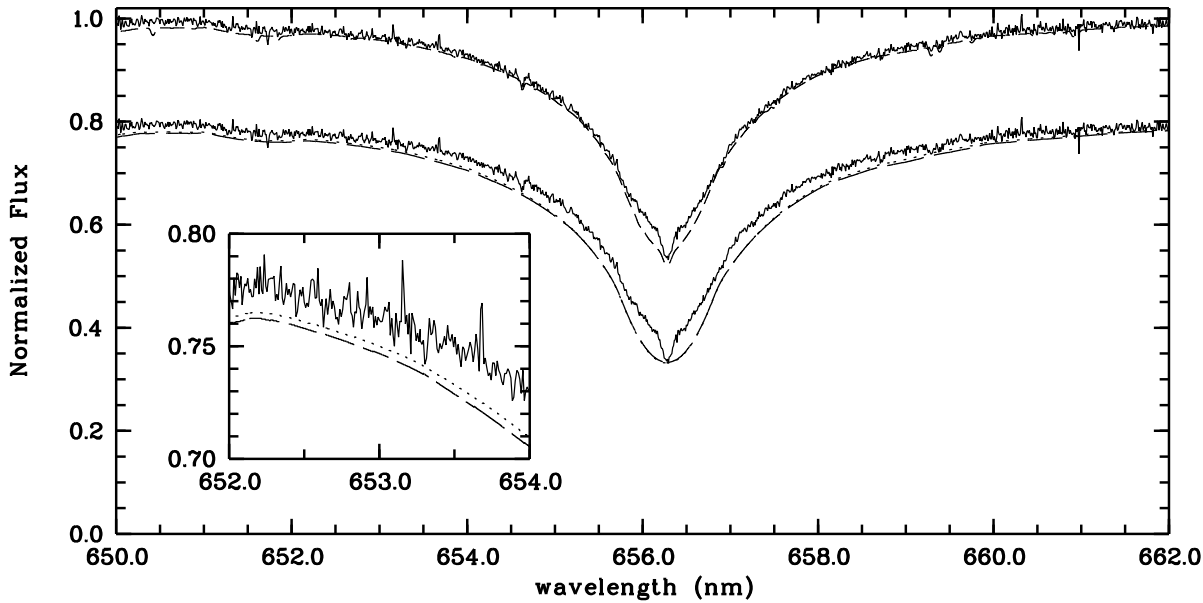,clip=true}
\caption{
The observed spectrum of HD 141851 (solid line) superimposed 
on the combined spectrum (short dashed line) obtained by combining 
two synthetic spectra as described in the text (upper panel).
In the lower panel the same observed spectrum is compared
with two synthetic spectra computed with solar abundances and parameters 
\teff and log~g derived from the photometry:
8080~K, 3.86 (long dashed line) and 8258~K, 3.69 (dotted line); 
the two computed spectra
are almost identical as can be seen in the inset.
}
\label{hd14185Ha}
\end{figure*}

\begin{figure}
\psfig{figure=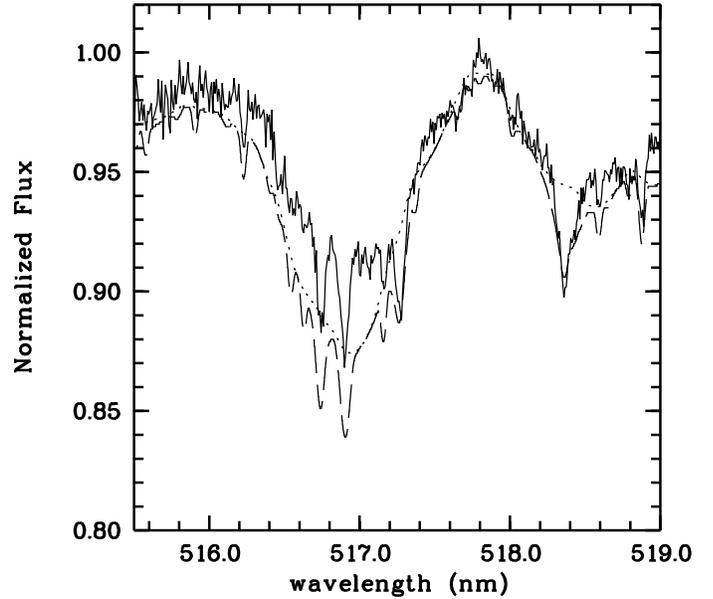,clip=true}
\caption{ The observed Mg I triplet of HD 141851 
compared with two synthetic spectra:
that computed with MD parameters (dotted line) and that obtained from the 
combination described in the text (long dashed line). 
}
\label{hd141851_mgI}
\end{figure}

We note that what seems  high noise of the spectrum
of this star
are, in reality,  many weak lines of a component
with a lower \teff and a much lower \vsini ($\leq$ 20 km s$^{-1}$).

According to our spectra, the lines of the secondary component are weak, 
but definitely present in the
blue region of the spectrum  and more clearly
from the Na I doublet.

The observed colour indices have been used in both photometries to
derive the atmospheric parameters; in fact E(b-y)=-0.010 (computed by 
the program of Moon, 1985).
This slightly negative colour index may be considered as due both to
the errors on the photometric measurements and to a distorted flux
due to the composite spectrum. Penprase (1993) gives
E(B-V)=0.07 which corresponds to E(b-y)=0.05. According to the distance
of HD 141851 obtained from Hipparcos  data, combined to the study 
of the high galactic latitude molecular clouds by Penprase (1993),
such a large reddening seems improbable. We remind that E(B-V) has been derived
by combining two photometric systems as well as the spectral classification
of the star (Penprase, 1992).

We computed two synthetic spectra with the parameters \teff
and log~g derived from uvby$\beta$ and Geneva photometric colour indices
as listed in Table~2,
namely, (8080~K, 3.86), and (8258~K, 3.69).
We assumed  v{\it sini} = 250 km s$^{-1}$ from Royer (2001) and used
solar abundances models.

In Fig. \ref{hd14185Ha} the observations of the H$_\alpha$  region are 
overimposed on these two computed spectra. This comparison
clearly shows:
\begin{enumerate}

\item the difference of \teff and log g computed by MD and K\"unzly et al. 
calibrations are such that the two different combinations produce the same
Balmer line profiles 

\item the flat core of the observed H$_\alpha$ (and H$_\gamma$) profile,
with respect to the computed one, is one of the binary signatures according 
to Faraggiana et al. (2001a), while the narrow core component 
belongs to the companion star.

\end{enumerate}

The aspect of the observed spectrum, the value of \teff obtained from the 
combined light, the magnitude difference in the H-band (Gerbaldi et al., 2003)
and the analysis 
by Grenier et al. (1999) allow us to derive, as a starting guess, a 
combination of two spectra having \teff= 8000 and 6000 K, log g = 4.0.
The luminosity ratio in V has been computed from the absolute magnitudes
M$_H$ and M$_V$ as given in the Allen's astrophysical quantities (Cox, 2000)  
for dwarf stars A7 and G0 
which correspond to the two \teff values. The M$_H$ values of these spectral 
types are coherent with the luminosity ratio observed in AO. For the V 
magnitude, it follows that the ratio of the luminosity in this spectral
domain is L$_1$/L$_{tot}$=0.84 and L$_2$/L$_{tot}$=0.16. This combined 
spectrum gives a better fit to the observation (Fig.   \ref{hd14185Ha} and 
\ref{hd141851_mgI}), but still
not in a satisfactory way.

We only  conclude that this is another star for which the claimed metal
deficiency cannot be investigated
if the duplicity of the system is not taken into account.

\subsection {HD 196821}

Various classifications of this star (classified as variable in 
the SIMBAD database)
appear in the literature: magnetic 
star according to Wolff \& Preston (1978) who measured
\vsini=20 $\mathrm{km\, s^{-1}}$;
Ap of the Si-Cr group according to Heacox (1979);
B9-HgMn in the Renson et al. (1991) catalogue of Ap stars, 
\lambo according to Abt \& Morrell (1995) who measured 
\vsini=10 $\mathrm{km\, s^{-1}}$.
A positive value of $\Delta$a (+0.014) is observed by Maitzen et al. 
(1998), contrary to what is expected in \lambo stars (Maitzen
\& Pavlovski, 1989).

\begin{figure}
\psfig{figure=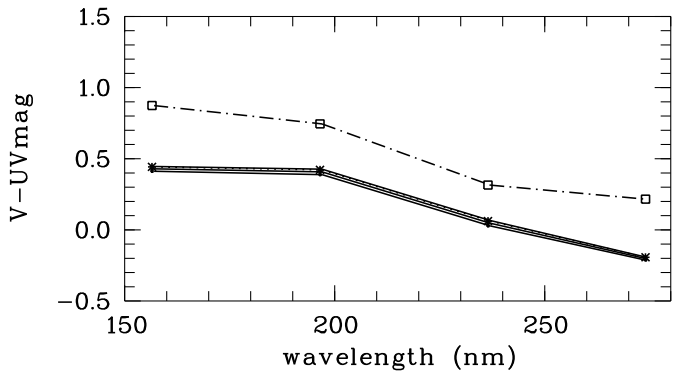,width=8.8cm,angle=-90,clip=true}
\caption{The UV flux (and associated error bars) of HD 196821 measured by TD1 
compared to the theoretical flux (dash-dot line) computed with 
solar abundances. }
\label{HD196821_TD1_m10}
\end{figure}

The high blanketing, characteristic of CP stars and opposite to what 
is expected 
in \lambo stars is the cause of the UV flux distribution shown in 
Fig. \ref{HD196821_TD1_m10}.
In fact the UV flux observed by TD1 indicates that the blanketing is higher 
than that predicted for a star with solar abundances, so excluding that 
HD 196821 is a \lambo star.

In this figure  the 4 observed magnitudes are derived
from the absolute fluxes in four passbands centered at 274.0, 236.5, 
196.5 and 156.5 nm. The computed magnitudes are derived from the theoretical
Kurucz fluxes integrated over the profile of the photometer channel passband 
(for the magnitude centered at 274.0) and over the response of each of the 
three spectrophotometer channels  (135.0-175.0, 175.0-215.0, 215.0-255.0). 
These profiles are given in the Thompson et al. (1978) catalogue.
The revised version, available at CDS,  of this catalogue has been used to
compute the UV magnitudes and their errors.

The only abundance analysis is that made by Heacox (1979) and it
is based on \teff= 10400 K, log g = 3.7, ${\xi}$=3 \kms and \vsini=20 
$\mathrm{km\, s^{-1}}$.

\begin{figure}
\psfig{figure=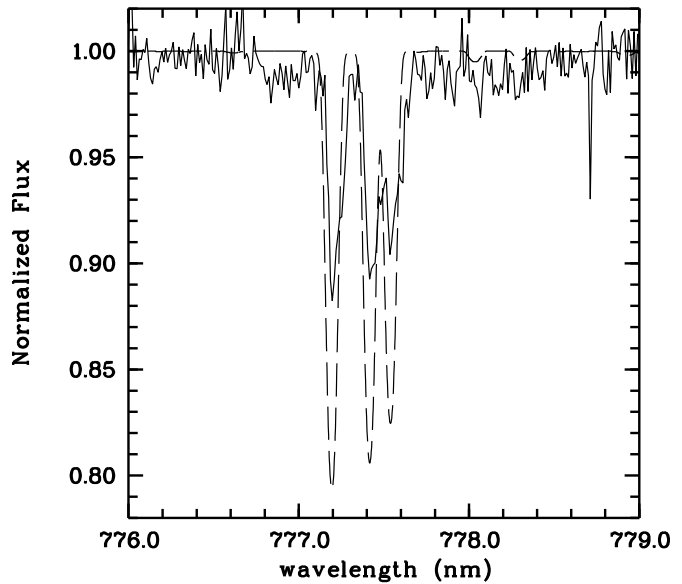,width=8.8cm,angle=0,clip=true}
\caption{The OI 777.4 triplet of HD 196821 compared to the theoretical spectrum
computed by assuming  solar abundances.}
\label{HD196821_oxy}
\end{figure}

In his classification Bidelman (1988):
notes "rather odd spectrum; Ap Cr?, SB?". 
In fact, the inspection of the spectrum 
taken on 2002, Oct 6 confirms that HD 196821 is an SB; the presence of a 
companion star is revealed by the same kind of asymmetry of the strong
lines; the profile  of the O I triplet is 
given in Fig. \ref{HD196821_oxy}
to illustrate the shape of the observed lines,
which should be  unblended in a single star. 

\subsection {HD 210111}

In their RV program Grenier et al. (1999) classified the star as probable
double.

\begin{figure}
\psfig{figure=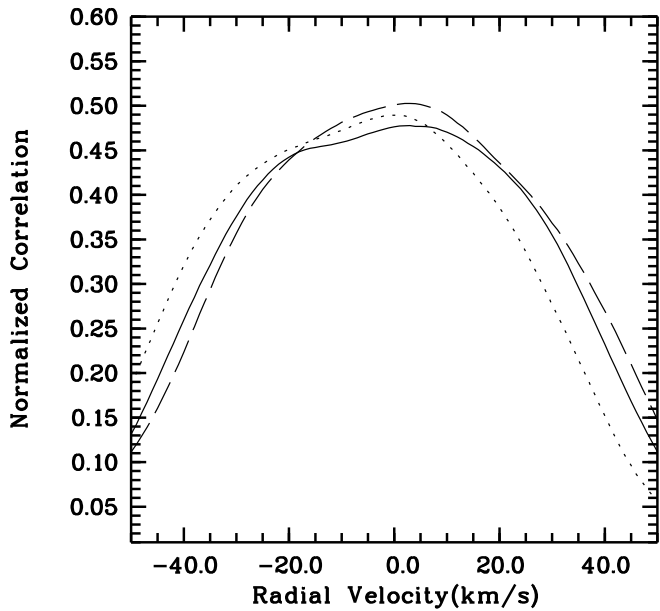,clip=true}
\caption{ The cross correlations of the three spectra of  HD 210111 [taken on
Sep 7th, 1993 (solid line), Sep. 18th, 1994 (long-dashed line) 
and on  Nov. 14th, 1994 (dotted line)] 
with a computed spectrum  (\teff = 7500 K, log g = 4.0 $v\sin i=5$ \kms)
over the range 420-450nm.  Both in the synthetic
and observed spectra the 
 H$_\gamma$ has been artificially suppressed (see text). 
}
\label{fig_reshd210111}
\end{figure}

The duplicity of this star is suggested by the asymmetric and distorted 
profiles shown in figures 1 and 3 of Holweger \& St\"urenburg (1991)
paper, especially for the NaI doublet, for the Mult. 4 Fe I lines 
392.026, 392.2914, 392.7922 and
393.0299 nm and for Al I 394.4009. 
The same double structure of Na I is evident in the
average spectrum  (of 5 spectra taken in 50 mn of observations) 
given by Bohlender et al. (1999) in their figure 9.
Bohlender et al. (1999) in their study of non-radial pulsation (NRP) 
of \lambo stars underline that this
object has the largest amplitude NRP ever observed in dwarf
stars of any class. 
Bohlender et al. report a peak to peak 
amplitude of 3\% of the continuum
in the mean-absolute-deviation spectrum (MAD).
These authors do not report any radial velocity variations,
which, as in all non radial modes, must be small.
One may estimate the amplitude of
radial velocity variation in \kms by multiplying by a factor
of 100 the amplitude in the MAD (Mantegazza, private
communication), we thus expect radial velocity
variations with an amplitude of the order of 3 \kms.
The multiperiodic high amplitude NRP is amply documented 
by these authors.
The cross-correlations computed with the 3 spectra obtained 
by us are
shown in Fig. \ref{fig_reshd210111}.
The asymmetry of our correlation
curves refers to spectra of 1h-2h of exposure; the  largest amplitude
of the NRP measured by Bohlender
has a period of 49 minutes, so the effect of NRP has been averaged
on these spectra. Nevertheless the RV between the two deformed peaks
of the cross correlation is more than 20 \kms, 
which cannot be confused with the radial velocity variations
induced by the NRP, which are an order of magnitude smaller.
Our results combined with Bohlender et al. (1999) remarks suggest that 
HD 210111 is  probably a combination of two similar NRP stars.

\begin{figure}
\psfig{figure=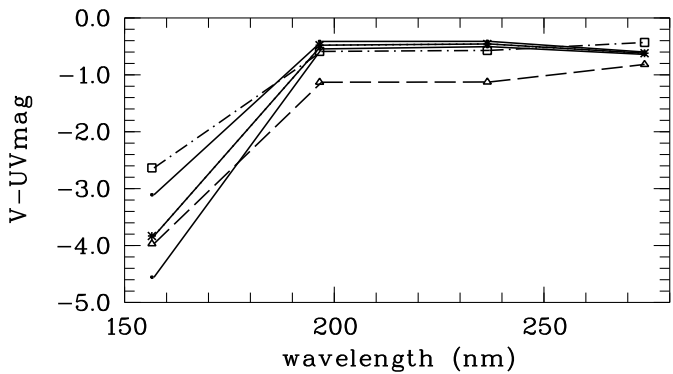,width=8.8cm,angle=-90,clip=true}
\caption{The UV flux of HD 210111 and associated error bars (thick lines) 
compared to the theoretical flux (\teff = 7500 K
log g = 4.00   
abundances 1 dex lower than solar) (dash-dot line) and solar abundances 
(dashed line).}
\label{HD210111_TD1_m10_1_revised}
\end{figure}

The TD1 observations suggest a flux deficiency in the band 156.5 nm
compared with the others 
(Fig. \ref{HD210111_TD1_m10_1_revised}) 
that, if confirmed, indicates that
the companion star has a lower \teff.

\bigskip

\subsection {HD 149303: wrong binary detection ?}

This star is a 
binary according to Paunzen et al. (1999a) from the profile of 
the O I triplet; one component being a low rotating star 
and the other one a fast 
rotator. In spite of this duplicity detection, the abundances of O, N, 
and S have
been determined by these authors and by and Kamp et al. 
(2001) in LTE and NLTE treating the star as a single one. 

The values of \teff (8000 K) and log g (3.8) used in these papers are 
those derived from
Str\"omgren photometry through the calibration by Napiwotzki
et al. (1993). The \vsini = 275 \kms has been derived also, 
in the hypothesis that the spectrum is that of a single object.

In another paper
Paunzen et al. (1999b) determined 
LTE abundances  of Mg, Ti, Fe and Ni using totally different values 
for \teff and log g (9000 K and 4.2) derived from another calibration of 
photometric data ( (b-y)-c$_1$ by Kurucz (1991)). The value of \vsini 
(200 $\mathrm{km\, s^{-1}}$) derived in this paper is also very 
different from the previous one. 
We note here that the (b-y) colour
is inconsistent with the ${\beta}$ value; in fact this latter
predicted from the other stellar colours is 2.875 instead of the 
observed value of 2.848.  
It does seem a bit odd that abundances of different elements
are derived using so different atmospheric parameters.

The \teff values that we have derived from the uvby${\beta}$  and Geneva 
colour indices are 8120 and 8483 K, and none of them agree with the 
H$_{\alpha}$ profile.
The star is the A component of a visual binary system; the E(b-y) computed
for its B companion is 0.031, while E(b-y)=-0.018 has been derived for the
primary. By forcing the star in another MD group (gr 5), the consistent 
E(b-y)= 0.023 is obtained for both; however the 
\teff computed in that case 
is 9520 K (log g = 3.85) 
which also does not fit the observed H$_{\alpha}$ profile
and is in strong 
contradiction with the UV flux observed by IUE.

\begin{figure}
\psfig{figure=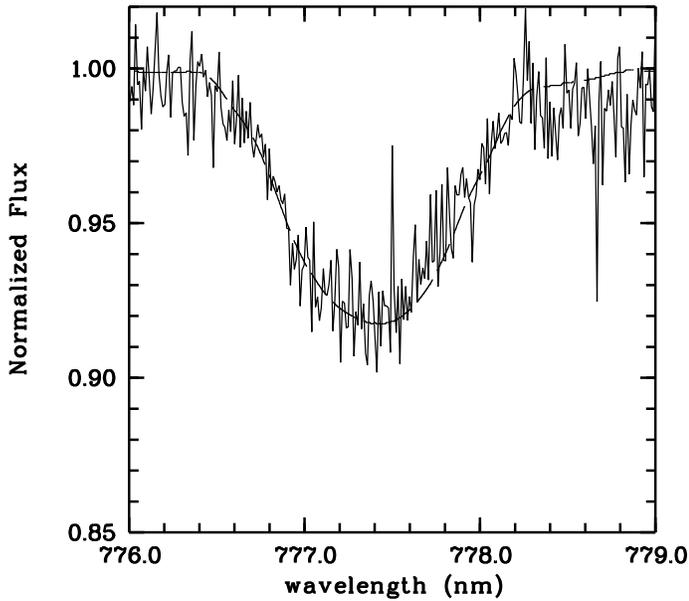,clip=true}
\caption{ The OI 777.4 profile of HD 149303 compared with that computed
in the hypothesis that the star is a single object rotating at 
275 $\mathrm{km\, s^{-1}}$;
the fictious fit has been obtained by adopting the MD parameters, but
increasing the oxygen abundance by 1 dex to simulate the NLTE effect on
this triplet}
\label{fig_hd149303oxy}
\end{figure}

The complex structure of the OI 777.4 triplet claimed by Paunzen et al. (1999a) 
is not present on our spectrum, which  simply indicates
a very fast rotator (Fig. \ref{fig_hd149303oxy}).

\begin{figure}
\psfig{figure=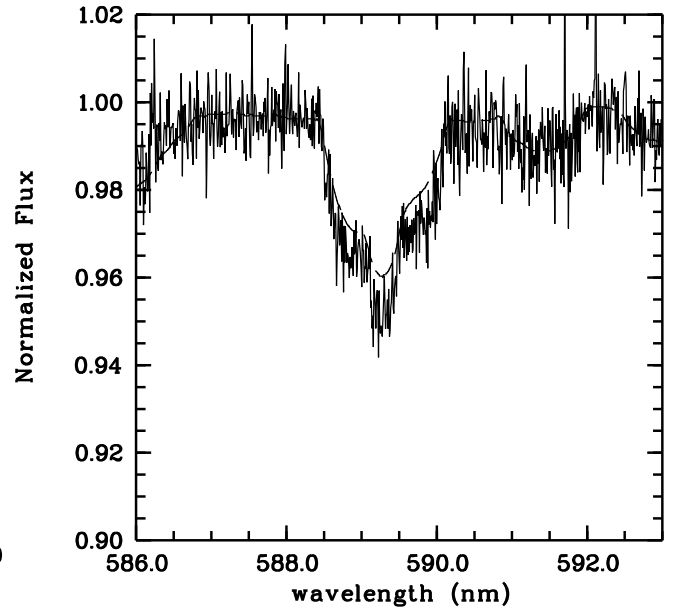,clip=true}
\caption{ The NaI doublet appears as a single very broad feature in the
spectrum of HD 149303 (thick line); the synthetic spectrum has been 
computed by assuming that the star is a single object rotating at 275 
$\mathrm{km\, s^{-1}}$.
The convolution of such broad lines produces the fictious
central peak.}
\label{fig_hd149303na}
\end{figure}

Even the Na I doublet (Fig. \ref{fig_hd149303na}) is merged in a single feature
with the central spurious "subrotational" peak due to the doublet 
convolution (see comments on this effect in Dravins et al., 1990). 

It might be interesting to follow this star to recover the phase at which
the low-rotating companion is detectable.

If this object, as one single star or as the fast rotating component of a 
binary system,
is really rotating at \vsini=275 $\mathrm{km\, s^{-1}}$, 
it should have lost its spherical shape and  the gravity-darkening effect 
cannot be neglected. 
The effects of such a rotation
on the stellar continuum and on EW of lines
should be analyzed. 
Clearly this analysis cannot be done on a  single spectrum  of low S/N ratio
such as the one at 
our disposal. We have only checked that the H$_{\alpha}$ profile is
not fitted by those computed by adopting the models with the parameters
derived by Str\"omgren and Geneva photometries; these models cannot be 
used for abundance analyses.
According to our 
interpretation, it is highly probable that the extremely high \vsini 
value is in reality a combination of those of two mean-fast rotating stars
whose average parameters do not reproduce the H$_{\alpha}$ profile because
the stars are not equal.

The Referee suggested that the visual  companion 
of HD 149303, whose separation has been
measured by Hipparcos (parallax=14.48 mas,
sep=16\farcs{36} , $\Delta$H$_p$=3.08) may have been in the spectrograph  
slit during the Paunzen
et al. observations and not during those presented here.
The presence of the low vsini component has been discovered by Paunzen 
et al. (1999a) on the OI triplet on spectra taken during 1996-1998 (no dates
are given for these observations). 
The epoch of the Hipparcos catalogue is 1991.25, thus
in order to be within the slit width of Paunzen et al. 
(which must have been of a few
arcseconds, at most)
the B component should have moved in less than 7 yrs at least 15  arcsec
to be confused with its brighter companion; its tangent velocity would
be  701.5 km/s.
Whichever 
the orbit inclination the system cannot be bound
and one should admit that the companion star
is moving on an hyperbolic orbit, this is highly 
unlikely and can be dismissed. Therefore the hypothesis that this
star was within the width of Paunzen's slit can be rejected. 
We have also checked on DSS II IR image taken on 1997.35 that the 
B companion was, at that date, at a distance of about 17 arcsec and therefore
could not be in the Paunzen et al. slit width.

An alternative hypothesis 
is that the companion star was within 
the height of the slit,  which  in fact may be quite long.  
However the spectra of the two stars
should have been clearly separated in the focal plane,
unless the instrument had a scale of $\sim 6''$per pixel. 
Therefore Paunzen et al. should have been able to extract
an uncontaminated spectrum of the primary star.
Thus the secondary star seen by Paunzen cannot be the one
seen by Hipparcos, but must be one which is much closer.

We recall also that
speckle measures by McAlister et al. (1997)  gave a negative result for
a nearby companion with
an upper limit for the separation  $\leq$ 0.038 arcsec.

Finally we recall that the known secondary with a magnitude difference of 3.08
corresponding to a flux ratio of less than 0.06 
can hardly affect the spectrum of the primary star.

\section {Discussion}

The very diverse properties  of the stars classified as \lambo
prompted us (Faraggiana \& Bonifacio , 1999) 
to consider the  hypothesis that 
the peculiarities in many, if not all, \lambo  stars
result from the fact that they
are unresolved binary or multiple systems
consisting of stars  of similar luminosity.
In this paper we have 
examined the extensive list of Gerbaldi et al. (2003) to
see how many of the 136 objects classified as
\lambo stars show composite spectra which,
if interpreted as spectra of a single star, may
be responsible of the peculiarities which led
to the \lambo classification.

Four stars in Gerbaldi et al. (2003) are misclassified 
beyond any reasonable doubt:
HD 34787, is a peculiar shell star with strong SrII 4077 and no shell core
of Balmer lines (Gray et al. (2001), HD 37886 is a Hg-Mn star, HD 89353
(HR4049) 
is an extremely iron deficient
post-AGB star, HD 108283 is a shell star.
Therefore 
the \lambo candidates to be considered 
are 132.

One has to bear in mind the difference between binary systems
in which one of the companions is several magnitudes
fainter than the other (as is the case for, e.g., many Am stars)
and binary systems which give rise to composite spectra.
Let us summarize all the stars for which  the presence
of a composite spectrum is either sure or highly probable.

\subsection{Stars classified as \lambo with composite spectra }

We consider as binaries which 
surely produce composite spectra those for which
a separation between the components of less than
2 arcsec has been 
measured with  a magnitude difference not larger than 2.2 mag. 
Twelve systems satisfy these criteria:
9 systems  detected by Hipparcos (HD 22470, HD 36496, HD 38545,
HD 47152, HD 97773, HD 118623, HD 170000, HD 217782 and HD 220278); 2 systems 
in the WDS (Washington Double Star) Catalog (Worley \& Douglass, 1997)
(HD 160928 and HD 290492); 1 star measured with adaptive optics (HD 141851)
(Gerbaldi et al., 2003).

In addition to the stars for which high angular resolution
observations have detected the duplicity and allowed to measure the separation,
we must consider those stars for which
a detailed inspection of high resolution spectra revealed the
composite spectrum:
HD 64491 (Faraggiana \& Gerbaldi, 2003), 
HD 111786 (Faraggiana et al., 2001a),
HD 153808 (Faraggiana et al., 2001a),
HD 174005 (Faraggiana et al., 2001b),
as well as 
the four stars discussed in the previous sections (HD 11413, HD 79108, HD 196821
and HD 210111).

To the list of stars with composite spectra
two known binary systems must be added: the SB2 HD 210418 (Gray \& Garrison
1987) and HD 98353; this latter star, better known as 55 UMa is
a triple system  formed by a close pair (A1V and A2V,
both moderately Am) and a more distant component (A1V); it is
discussed in detail by Liu et al. (1997).

It thus appears that 
the percentage of binaries in
our sample,  whose angular separation and magnitude difference
produce composite spectra because not 
resolved on a spectrograph entrance, is 22 out of 132, i.e.
of the order of 17\%, therefore
not as small as assumed by people discussing the origin of these
stars.

\subsection{Stars classified as \lambo which likely display
composite spectra }

The above percentage could be higher, if a few
of the below listed
stars, for which a composite spectrum can be suspected,
can be confirmed as composite spectrum stars.

Stars for which speckle interferometry measured a separation less than 0.1
arcsec, but for which there are no indications on the magnitude difference
(HD 21335, HD 98353, HD 225218) may be suspected
of producing composite spectra, but  require complementary observations  to
assess whether their spectra are composite indeed.
HD 98353=55 UMa is a case for which such
observations are available and it is in fact  the 
triple system discussed above.

Other binary candidates which may produce composite spectra are the 14 stars
with variable RV,  not belonging to known visual or spectroscopic
binary systems, already listed in Gerbaldi et al. (2003):
HD 39283, HD 56405, HD 74873, HD 79108, HD 87696, HD 111604, HD 125489,
HD 138527, HD 169009, HD 177756, HD 179791, HD 183324, HD 220061, HD 221756.
A faint companion of HD 138527 has been detected by
adaptive optics (Gerbaldi et al., 2003).
Although other phenomena, such as, e.g., stellar pulsations,
may be responsible for variable radial velocities,
it is likely that several of these turn out to be 
spectroscopic binaries.
In fact among these 
HD 79108 is the new multiple system discussed in section 3.2.
For two others, HD 56405 and HD 125489,  the UV flux  (TD1 UV magnitudes),
is lower than that computed with solar abundances, contrary to what is
expected in the low-blanketed \lambo stars (see Gerbaldi et al., 2003).
We   consider this 
an indication
of composite spectra. In fact 5 out of the 8 stars showing this low UV flux
are already known to be close binaries producing composite spectra, the 
remaining three are the two above stars and HD 212150.
For 4 of the variable RV stars
the UV flux is better fitted by
theoretical fluxes computed assuming
a  solar abundance than lower than solar
abundances, as should be the case for \lambo stars 
(HD 39283, HD 74873,
HD 87696 and HD 179791). Again this is a reason
to suspect a composite spectrum.

A further reason to suspect the existence
of composite spectra
is the discrepancy between the absolute magnitude 
derived from the measured parallax and that 
derived from some calibration of photometric
indices. 
From the comparison
between the absolute
magnitude derived from the Hipparcos parallax
 and the calibrations adopted by Moon (1985), 
a slight systematic difference esists,
in the sense that, on average, the Mv derived from
the Hipparcos parallax is brighter than 
that derived from the Moon (1985) calibrations.
It is probably significant that the majority of
the known binary stars appear over-luminous
with respect to what expected from colours
( see figure 4 in Gerbaldi et al. (2003)).
This is not always the case, for some
stars the Hipparcos absolute magnitude is
in excellent agreement with that derived from
the Moon (1985) calibrations  and in some others it is
even slightly fainter. Therefore the
criterion of over-luminosity 
is neither necessary nor sufficient 
to establish the multiple nature of  a star.
However it  certainly  prompts for further 
observations to ascertain its nature.

We add also that 16 \lambo stars are included in the
Grenier et al. (1999) catalogue of a sample of 610 southern
B8-F2 stars. Radial velocities are measured using a 
cross-correlation method (the templates are a grid of 
synthetic spectra).  For 2 of them (HD 183324 and HD 223352) 
no RV is given because the
cross correlation  with the adopted template produced a 
correlation peak too low and asymmetric; 2 other
(HD 111786 and HD 141851) are 
certain doubles, 4 (HD 30422, HD 31295, HD 193281 and HD 210111)
are probable and other 4 (HD 319, HD 75654, HD 79108 and HD 204041) 
suspected doubles, 1 (HD 142703) is a probable multiple system, 1
(HD 170680) showed the
wide peak of B stars and only 1 (HD 39421) had symmetric and 
gaussian profile.
We cannot consider these remarks as a definite proof of 
the duplicity of these stars, but an indication that
a careful check for non-duplicity is required before
to elaborate abundance analyses of them. The fact that 14 over 
the 16 examined \lambo stars gave problems in the RV measures 
is a further indication of the high percentage of composite spectra
objects among \lambo stars.

We finally note that for 19 stars 
the UV flux (TD1)  cannot be reproduced by
any LTE computation (groups 3a and 3b of Gerbaldi et al., 2003).
Five of these stars are already known to produce composite
spectra, two are classified D and one U by Hipparcos,
while six have not been observed by this satellite.
A distorted UV flux is not by itself evidence
of duplicity, however alternative explanations are 
more contrived. 
Therefore  these stars ought to
be further scrutinized in order to understand the
reasons for this anomalous flux distribution.

\section{Conclusions: duplicity and \lambo nature}

From the above discussion we may conclude
that the percentage of \lambo stars
with composite spectra is at least 17\%, but could 
rise up to 28\% if all the stars for which 
a composite spectrum is suspected were confirmed,
and even more if we consider that for several  stars,
mainly the faintest ones, there is no data beyond classification
dispersion spectra.

\balance
In order to understand the \lambo phenomenon
the list of \lambo stars should
be cleared of all the above stars
for which the \lambo classifications has
not been confirmed when taking into account the
duplicity.
For example 
for the SB2 HD 84948 we have shown that the current
\lambo classification  of the
two components is based on an abundance analysis made by assuming incorrect
atmospheric parameters (Faraggiana et al., 2001a). Therefore the \lambo status
of this star has still to be demonstrated.

The claim of Faraggiana \& Bonifacio (1999) that 
undetected duplicity often results in {\em underabundances},
when the spectrum is subject to a standard abundance analysis
has recently found an independent confirmation
by the analysis of overluminous F-type stars by
Griffin \& Suchkov (2003).
These authors analyzed F stars for which the absolute magnitude
derived from uvby$\beta$ photometry is over  0.5 mag greater
than that derived from the Hipparcos parallax.
In their sample of 77 stars they found 27 new binary systems
from the radial velocity survey.
The percentage of newly detected SB2 systems with positive [Fe/H] is 6 \%, while
the same percentage on the whole sample is 21 \%, clearly
showing that a composite spectrum  may simulate
metal underabundance more often than not.

At present the connection between the \lambo nature and
duplicity, if any, is not clear.
The evidence we have presented here implies
that a non-negligible fraction of \lambo stars  has
composite spectra due to duplicity.
Furthermore the effect of duplicity, when neglected,
is such as to mimic underabundances, which is one of
the main characteristics of \lambo stars.
Therefore for all the stars 
with composite spectra the \lambo classification needs
to be re-established (or rejected) taking into 
account the composite nature of the spectra. 
It would be tempting to conclude that the \lambo
phenomenon is {\em solely} due to undetected duplicity.
However, on the one hand, there exists a majority of stars
for which no sign of duplicity has been detected, although 
one should keep in mind that 
in some cases duplicity has not been searched for, or 
$v\sin i$ is too large for a spectroscopic detection or
the stars are too faint and the angular separation too small
to be resolved visually.   On the
other hand, it may well be that the \lambo stars with
composite spectra, when analyzed taking into account
their binary or multiple nature, still show
the underabundances typical of \lambo stars.

If a significant number of \lambo stars with composite
spectra, when properly analysed, 
show ``normal'' (i.e. solar or solar--scaled) chemical
abundances the very existence of \lambo stars
as a class of chemically peculiar stars may be called into
question.

We believe that the most urgent thing to do, at present,
in the study of \lambo stars is
to perform such a detailed analysis for as many 
\lambo stars with composite spectra as possible.
It is of paramount importance that such analysis 
is based on many spectra, taken at different phases,
with as large a spectral coverage as possible.

The other way to tackle the problem is to
pursue observations at high angular resolution, 
interferometric instrumentation on
large telescopes, especially to VLTI should add valuable
information, allowing
to explore  effectively 
the problem of the \lambo stars. 
If the \lambo phenomenon is due to 
undetected duplicity one may expect
the detection of new previously unknown 
binary systems.

What we now consider firmly established is:

\begin{enumerate}

\item the analysis of the \lambo binaries abundances {\bf must} be done by
disentangling the observed composite spectra;

\item in order to know 
the position of the stars of a multiple
system in the HR diagram one has to know the \teff and L of all the
components.  Adopting average values of both 
parameters to speculate on the evolutionary
stage of \lambo stars does not have any physical meaning.

\end{enumerate}

This paper concludes our series  on the \lambo candidates.
We believe we have convincingly demonstrated
the need for high resolution spectroscopy over a large
wavelength interval and spectrum synthesis in order to determine 
the chemical composition of the \lambo candidates with  composite
spectra. Such an analysis has not been performed for
any of them, yet.
The current generation of echelle spectrographs on 4m and 8m class
telescopes is well suited to tackle the problem.

\begin{acknowledgements} Use has been made of the SIMBAD database operated
at the CDS, Strasbourg, France.
We are grateful to 
Fiorella Castelli  
for  many helpful discussions on
model atmospheres and spectrum synthesis 
and to Luciano Mantegazza for illuminating discussions
on $\delta$ Scuti and other pulsating stars. 
This research was done with support from the Italian MIUR  
COFIN 2002028935-003 grant.

\end{acknowledgements}

\end{document}